\documentclass[superscriptaddress,showpacs,twocolumn]{revtex4}
\usepackage{amssymb}
\usepackage[tbtags]{amsmath}
\usepackage{graphicx}
\usepackage{epsfig,graphicx,times}

\begin{document}
\date{\today}
\title{Effects of dynamical phases in Shor's factoring algorithm with operational delays}
\author{L.F. Wei}
\affiliation{Frontier Research System, The Institute of Physical
and Chemical Research (RIKEN), Wako-shi, Saitama, 351-0198, Japan}
\affiliation{Institute of Quantum Optics and Quantum Information,
Department of Physics, Shanghai Jiaotong University, Shanghai
200030, P.R. China}
\author{Xiao Li}
\affiliation{Department of Physics, Shanghai Jiaotong University,
Shanghai 200030, P.R. China} \affiliation{Department of Physics,
PMB 179, 104 Davey Laboratory, Penn State University, University
Park, PA 16802-6300, USA}
\author{Xuedong Hu}
\affiliation{Frontier Research System, The Institute of Physical and Chemical
Research
(RIKEN), Wako-shi, Saitama, 351-0198, Japan}
\affiliation{Department of Physics, University at Buffalo, SUNY, Buffalo
NY14260-1500, USA}
\author{Franco Nori}
%\thanks{corresponding author}
\affiliation{Frontier Research System, The Institute of Physical
and Chemical Research (RIKEN), Wako-shi, Saitama, 351-0198, Japan}
\affiliation{Center of Theoretical Physics, Physics Department,
Center for the Study of Complex Systems, University of Michigan,
Ann Arbor, Michigan 48109-1120, USA}
\thanks{Permanent address}

\begin{abstract}
\vspace{-0.2cm} Ideal quantum algorithms usually assume that
quantum computing is performed continuously by a sequence of
unitary transformations. However, there always exist idle finite
time intervals between consecutive operations in a realistic
quantum computing process. During these delays, coherent ``errors"
will accumulate from the dynamical phases of the superposed wave
functions. Here we explore the sensitivity of Shor's quantum
factoring algorithm to such errors. Our results clearly show a
severe sensitivity of Shor's factorization algorithm to the
presence of delay times between successive unitary
transformations. Specifically, in the presence of these {\it
coherent ``errors"}, the probability of obtaining the correct
answer decreases exponentially with the number of qubits of the
work register. A particularly simple phase-matching approach is
proposed in this paper to {\it avoid} or suppress these {\it
coherent errors} when using Shor's algorithm to factorize
integers. The robustness of this phase-matching condition is
evaluated analytically or numerically for the factorization of
several integers: $4,\,15,\,21$, and $33$.
\end{abstract}

\pacs{03.67.Lx}
\maketitle

\vspace{-0.2cm}

\section{Introduction}

Building a practical quantum information processor has attracted
considerable interest during the past decade \cite{Bennett00}.
With the resources provided by quantum mechanics, such as
superposition and entanglement, a quantum computer could achieve a
significant speedup for certain computational tasks. The most
prominent example is Shor's factoring algorithm
\cite{Shor94,Vandersypen01}, which allows an exponential speedup
over the known classical algorithms.
The proposed quantum algorithms are constructed assuming that all
quantum operations can be performed precisely. In reality, any
physical realization of such a computing process must treat
various errors arising from various noise and imperfections (see,
e.g., \cite{Mussinger00}). Physically, these errors can be
distinguished into two different kinds: incoherent and coherent
errors. The incoherent errors originate from the coupling of the
quantum information processor to an uncontrollable external
environment, which is stochastic, and results in decoherence.
Coherent errors usually arise from non-ideal quantum gates which
lead to unitary but non-ideal temporal evolutions of a quantum
computer.
So far, most previous works (see, e.g.,
\cite{Shor95,Miquel96,Lidar98,Long,Plenio97}) have been concerned
with quantum errors arising from the decoherence due to
interactions with the external environment and external
operational imperfections. Here, we focus instead on internal
ones. The coherent errors we consider here are related to the {\it
intrinsic} dynamical evolution of the qubits {\it between}
operations.

A quantum computing process generally consists of a sequence of
quantum unitary operations. These transformations are usually
applied to the superposition states so that the quantum computer
evolves from an input initial state to the desired final state.
If the two qubit levels have different energies, as it is usually
the case, the superposition wave function of the quantum register
undergoes fast coherent oscillations during the finite time delay
between two consecutive operations. These oscillation, if not
controlled, can spoil the correct computational results expected
from the ideal quantum algorithms, where operational delays are
neglected.

In principle, these coherent errors can be either (1) avoided by
tuning the relevant energy splittings of the qubits to zero
\cite{MSS99,Feng01}; or (2) eliminated by introducing a ``natural"
phase induced by using a stable continuous reference oscillation
for each quantum transition in the computing process
\cite{Berman00}. Experimentally, for example in NMR systems (see,
e.g., \cite{Vandersypen01,Chuang97}), these errors were usually
corrected by introducing two additional operations before and
after the delay to reverse each undesired free evolution.

In this paper we perform a quantitative assessment of the effects
of the dynamical phases in Shor's algorithm by realistically
assuming that operational delays, between successive unitary
transformations, exist throughout the computation. We explore a
phase-matching approach to deal with the dynamical phase problem.
We show that coherent ``errors'' due to these phases, acquired by
the dynamical evolution of the superposed wavefunction during the
operational delays, may be avoided by properly setting the
\textit{total} delay. We then carefully evaluate the robustness of
such a phase-matching condition, focusing on its dependence on the
number of qubits, the length of the delay, and the fluctuations in
the qubit energy splitting.  Our discussions are in the context of
Shor's algorithm, but can be extended to other quantum algorithms,
such as the phase estimation and other algorithms~\cite{Wei04-2}.
For simplicity and clarity, here we assume that the influence of
the environmental decoherence and the gate imperfections on the
computing process are negligible.

The paper is organized as follows. In Section II, we present a
decomposition of Shor's algorithm and explain how we incorporate
the dynamical phases into the realization of this algorithm. The
usual decompositions of quantum algorithms into consecutive
elementary gates are strictly limited by the short decoherence
time. Here, we reconstruct the standard Shor's algorithm out of
four functional unitary transformations, and only consider the
operational delays between these larger building blocks. We assume
that each block can be exactly performed by only one-time
evolution as a multi-qubit gate (see,
e.g.,~\cite{wei04-1,Niskanen03}), avoiding the existing idle time
inside it. It is shown that the effects of dynamical phases are
not negligible, even in this primary or ``coarse-grained"
decomposition.
In Section III, we numerically evaluate several examples to
illustrate the phase-matching condition, and establish a clear
relationship between this condition and the equivalence of the
Schr\"{o}dinger picture and the interaction picture description of
a physical system.  We also demonstrate the robustness of the
phase-matching condition by varying the number of qubits involved,
the delay duration, and distribution of qubit energy splitting.
Finally, in section IV we present some conclusions and discussions
from our numerical studies.

\section{Four-block decomposition of Shor's algorithm with operational delays}

We study the dynamical phase problem in the context of Shor's
factoring algorithm.  In Shor's algorithm \cite{Shor94}, the
factorization of a given number $N$ is based on calculating the
period of the function $f(x)=a^x\, \mathrm{mod}\, N$ quantum
mechanically for a randomly selected number $a$ ($1 < a < N$)
coprime with $N$.  Here $y\,\mathrm{mod}\,N$ is the remainder when
$y$ is divided by $N$.  The order $r$ of $a\, \mathrm{mod}\, N$ is
the smallest integer $r$ such that $a^r\,\mathrm{mod}\,N=1$. Once
$r$ is known, factors of $N$ are obtained by calculating the
greatest common divisor of $N$ and $y^{r/2}\pm 1$.
A quantum computer can find $r$ efficiently by a series of quantum
operations on two quantum registers $W$ and $A$.  One is the work register $W$
with $L$ qubits, in which the job of finding the order is done; while the
values of the function $f(x)$ are stored in the auxiliary register $A$ with
$L^{\prime}$ qubits.  The sizes of the work and auxiliary registers are
chosen as the integers satisfying the inequalities $N^2 < q=2^L <2N^2$ and
$2^{L^{\prime}-1} < N < 2^{L^{\prime}}$.  Here $q$ is the Hilbert space
dimension of the work register.

As shown in figure 1,
\begin{figure}
\label{fig1}
\begin{center}
\vspace{1.8cm}
\includegraphics[width=12.8cm]{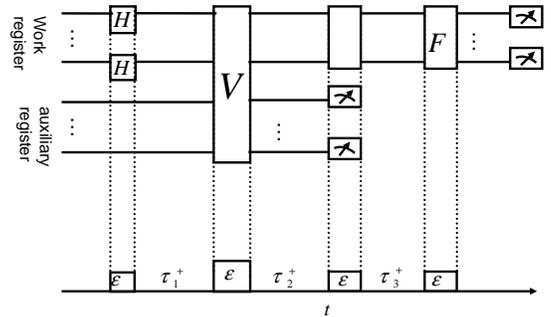}
\vspace{-6.6cm} \caption{Quantum circuit for implementing Shor's
algorithm with time delays $\tau_j^+\,(j=1,2,3)$ between the
successive operations. Here $H$ refers to a Hadamard gate, while
$F$ refers to a quantum Fourier transformation. Each block
operation is assumed to be exactly performed in a very short time
interval $\varepsilon$ (so that phases accumulated during the
operations are either accounted for by the operations themselves
or simply neglected).}
\end{center}
\end{figure}
a realistic implementation of Shor's algorithm can be decomposed
into the following unitary transformations:

1) Initialize the work register in an equal-weight superposition
of all the logical states, and the auxiliary register in its
logical ground state $|0\rangle_A$. Initially, each work qubit is
in its logical ground state $|0\rangle$. Assuming that a Hadamard
gate $H$ is applied to each qubit in the work register at one
time, the computational initial state of the system becomes:
\[
|\Psi(0)\rangle=\frac{1}{\sqrt{q}}\sum_{j=0}^{q-1}|j\rangle_W
\otimes |0\rangle_A.
\]
Here, the subindex $W$ stands for the work register state, and the subindex $%
A$ for the auxiliary register.
After a finite time delay $\tau_1$, and right before the second
unitary transformation is applied, the initial state
$|\Psi(0)\rangle$ of the whole system evolves to
\begin{equation}
|\Psi (\tau_1) \rangle = \frac{1}{\sqrt{q}}\sum_{j=0}^{q-1}
\,e^{-iE_j\tau_1}\, |j\rangle_W \otimes e^{-iE_0 \tau_1}\,
|0\rangle_A,
\end{equation}
with $E_j$ being the energy of state $|j\rangle$ and $\hbar=1$.
Here, $\tau_m\, (m=1,2,3,...)$ denotes the time interval between
the $(m-1)$th and $m$th unitary operations.
$\tau_m^+=\tau_m\,+\,\epsilon$ with $\epsilon\ll\tau_m$ being the
operational time of the $m$th unitary transformation, here assumed
to be extremely small compared to other time scales. In other
words, $\tau_m^+$ refers to the time interval between the end of
$(m-1)$th operation and the end of the $m$th operation. In what
follows, the global dynamical phase $\exp(-iE_0\tau_1)$ will be
omitted as it does not have any physical meaning.

2) Calculate the function $f_{N,a}(j)=a^j\,\mathrm{mod}\,N$ and
then entangle the work $\{|j\rangle_W\}$ and auxiliary registers
$|f_{a,N}(s)\rangle_A$ by applying a joint operation $\hat{V}$.
After another finite-time delay $\tau_2$ before the next step
(i.e., the third unitary transformation), the entangled state of
the whole system becomes
\begin{equation}
|\Psi(\tau_1^{+}+\tau_2)\rangle=\frac{1}{\sqrt{q}} \sum_{s=0}^{r-1}\,|\psi
\rangle_W \otimes|\phi\rangle_A,
\end{equation}
where
\[
|\phi\rangle_A=\exp[-iE_{f_{a,N}(s)}\tau_2]\,\,|f_{a,N}(s)\rangle_A,
\]
and
\[
|\psi\rangle_W = \sum_{l=0}^{w}\exp[-iE_{(lr+s)}(\tau_1^{+} +
\tau_2)]\,\,|lr + s\rangle _W,
\]
with $w=[(q-s-1)/r]$ being the largest integer less than
$(q-s-1)/r$. The dynamical phases of the qubits in the work
register, before and after the joint operation $\hat{V}$, can be
added directly, as this operator is diagonal in the logical basis.

3) Measure the auxiliary register $|\phi\rangle_A$ in its
computational basis $\{|j\rangle_A\}$. After this operation, the
state of the whole system becomes
$|\Psi(\tau_1^++\tau_2^+)\rangle=|\psi(\tau_1^++\tau_2^+)\rangle_W\otimes|
\phi(\tau_1^++\tau_2^+)\rangle_A$. In other words, the work and
auxiliary registers disentangle and the work register collapses to
one of its periodic states $|\psi(\tau_1^++\tau_2^+)\rangle_W$.

For example, if the measurement on the auxiliary register $|\phi\rangle_A$
gives a value $A_s=a^s\,\mathrm{mod} \,N $, then the work register
immediately becomes
\begin{widetext}
\[
|\psi(\tau_1^{+}+\tau_2^{+})\rangle_W =
\frac{1}{\sqrt{w+1}}\;\sum_{l=0}^{w} \exp[-iE_{(lr+s)}
(\tau_1^++\tau_2^+)]\,\,|lr+s\rangle_W.
\]
\end{widetext}
After the third unitary transformation is
applied, there is a third time delay $\tau_3$. The state
$|\psi(\tau_1^{+}+\tau_2^{+})\rangle_W$ now evolves to
\begin{widetext}
\begin{equation}
|\psi(\tau^{+}_1+\tau^{+}_2+\tau_3)\rangle_W=\frac{1}{\sqrt{w+1}}\;
\sum_{l=0}^{w}\exp\left[-iE_{(lr+s)}(\tau^{+}_1+\tau^{+}_2+\tau_3)\right]
\,\,|lr+s\rangle_W.
\end{equation}
\end{widetext}
Because of the collapse of the wavefunction $|\Psi(\tau_1^++\tau_2)\rangle$ in
Eq.~(2), the dynamical phases accumulated by the wavefunction
$|\phi\rangle_A$ of the auxiliary register do not affect the
algorithm anymore, as the relevant phase
$\exp[-iE_{f_{a,N}(s)}\,\tau_2^+]$ becomes a global phase.

4) Perform the fourth unitary transformation: the quantum Fourier
transform ($F$-Transformation) on the work register
$|\psi\rangle_W$, so that information regarding the order $r$ of
$a\,{\rm mod}\,N$ (i.e., the smallest integer $r$ such that
$a^r\,{\rm mod}\,N=1$) can be more easily extracted. After the
$F$-Transformation the state of the work register becomes
\[
|\psi(\tau)\rangle_W = \frac{1}{\sqrt{q}}\;
\sum_{k=0}^{q-1}\,g(k)\,|k\rangle_W\; , \,
\]
with
\[
\tau=\tau_1^{+}+\tau_2^{+}+\tau_3^{+}, \] being the time after
applying the fourth unitary transformation, and
\[ g(k)=\frac{\exp(2\pi
isk/q)}{\sqrt{w+1}}\sum_{l=0}^{w}\;\exp\left[-iE_{(lr+s)}
\tau+2\pi ilk\frac{l}{q}\right]\, .
\]
After another delay time $\tau_4$, i.e., right before applying the
fifth unitary transformation, the work register evolves into
\begin{equation}
|\psi(\tau+\tau_4)\rangle_W=\frac{1}{\sqrt{q}}\;
\sum_{k=0}^{q-1}\,g(k)\,e^{-iE_k\tau_4}\,|k\rangle_W.
\end{equation}

5) Finally, we carry out a measurement on the work register in the
computational basis $\{|j\rangle_W\}$ and derive the desired order $r$
satisfying the condition $a^r\,\mathrm{mod}\, N\,=\,1$. This measurement
yields the state $|k\rangle_W$ with probability
%
%\begin{widetext}
\begin{equation}
P(k)=\frac{1}{q\,(w+1)} \left|\; \sum_{l=0}^{w}\,\exp \left[
-iE_{(lr+s)}\tau+2\pi ilk\frac{r}{q} \right]\; \right|^2,
\label{eq-prob}
\end{equation}
%\end{widetext}
%
which is independent of the free evolution during the last delay
$\tau_4$. Notice that here $P(k)$ only depends on the {\it total}
effective delay time $\tau=\tau_1^+ + \tau_2^+ + \tau_3^+$, but
{\it not} directly on the {\it individual} time intervals
$\tau_m,\,m=1,2,3,4$.

In this decomposition of Shor's algorithm we have included time
delays only in between the various unitary operations, which were
implemented by independently using various one-time
evolutions~\cite{wei04-1,Niskanen03}. Note that only the delays
from the initial Hadamard gates to the finishing Fourier
transformation may result in physical effects. Fortunately, all
the operators during these delays are either diagonal or at least
not affecting the phase accumulation. Therefore, the phases in
each qubit simply add up.

If each unitary transformation is itself composed of several
consecutive steps, with delays between these internal steps, we
assume these delays to be negligible. This condition implies that
the internal time delays occurring between steps within each
unitary operation should be so short that their accumulated phases
are negligible. Such a condition is possibly difficult to satisfy
experimentally. However, our results below show that even under
such a restrictive condition the interference effects due to
dynamical phases between successive unitary transformation are
already too significant to be ignored.

For the ideal situation without any delay ($\tau_m\equiv 0$), the
probability distribution $P(k)$ in Eq. (5) reduces to that in the
original Shor's algorithm \cite{Shor94}. However, Eq.~(5) clearly
shows that the expected probabilistic distribution may be strongly
modified by the interferences due to the dynamical phases of the
superposition wavefunction, which would consequently lead to a
lower probability for obtaining the desired final output.

\section{Effects of dynamical phases}

In order to study the effects of dynamical phases, we need to
compute $P(k)$. The probability $P(k)$ in (5) can be computed if
the energies $E_{(lr+s)}$ for the various states $|lr+s\rangle_W$
involved are known exactly. These will be computed below.

\subsection{Phase-matching condition for eliminating the coherent errors due to operational delays}

  As a first approximation we assume that all qubits
in a quantum computer system possess identical energy spectra.
Such an approximation is valid for naturally identical systems
like trapped ions.  In this ideal case, when all the qubits have
the same energy splitting between ground and excited states,
different quantum states with the same number of excited qubits
will acquire the same dynamical phase.  For example, the
four-qubit states $|1_30_20_10_0\rangle$ and
$|0_30_20_11_0\rangle$ would acquire the same dynamical phase
$\exp(-i3\epsilon_0 t-i\epsilon_1 t)$ during a delay time $t$.
Here $\epsilon_0$ and $\epsilon_1$ are the energies of a single
qubit corresponding to the ground state $|0\rangle$ and the
excited state $|1\rangle$, respectively. Under this approximation,
equation~(\ref{eq-prob}) can be rewritten as
\begin{widetext}
\begin{equation}
P(k)=\frac{1}{q\,(w+1)} \left|\; \sum_{l=0}^{w}\,\exp \left[ 2\pi
ilk\frac{r}{q}
\right]\,\exp[-i(L^{(1)}_l-L^{(0)}_l)\tau\,\Delta]\;
\right|^2,\,\,\,\Delta=\epsilon_1-\epsilon_0,
\label{eq-prob1}
\end{equation}
\end{widetext}
where $\Delta$ is the qubit energy splitting and $L^{(1)}_l$
($L^{(0)}_l$) is the number of qubits in the logical state
$|1\rangle$ ($|0\rangle$) for the number state $|lr+s\rangle_W$.
Obviously, when the {\it total} effective delay time
$\tau$ ($\tau=\tau_1^++\tau_2^++\tau_3^+$) satisfies the
phase-matching condition
\begin{equation}
\tau\, \Delta=(\epsilon_1-\epsilon_0)\,\tau=2\,n\,\pi,\ \ \
n=1,2,3,..., \label{eq-phase}
\end{equation}
the above probability distribution $P(k)$ reduces to that of an
ideal computation process with $\tau\Delta=0$. This implies that
{\it the interference due to the fast evolution of the dynamical
phases can be suppressed periodically} so that the correct results
are obtained at the delay points indicated in (\ref{eq-phase}).

Physically, this phase matching condition is related to the
transformation of wavefunction from the interaction to the
Schr\"odinger pictures.  Theoretical derivations (see, e.g.,
\cite{CZ95}) for realizing quantum computation are usually in the
interaction picture, in which the Hamiltonian for the qubit
free-evolution does not appear, and the oscillation of the
superposed wavefunction does not exist.  More specifically, if a
system Hamiltonian $\hat{H}$ can be written as a sum of a free
oscillator part and an interaction part $\hat{H} = \hat{H}_0
+\hat{V}$, so that the time-dependent Schr\"{o}dinger equation can
be written as (in the so-called Schr\"{o}dinger picture where
operators are time-independent while states evolve with time)
$$i\hbar\frac{\partial}{\partial t} |\psi_S(t)\rangle = (\hat{H}_0 + \hat{V})
|\psi_S(t)\rangle \,,$$
one can introduce the interaction picture wavefunction
$|\psi_S(t)\rangle = \exp(-i\hat{H}_0 t/\hbar)|\psi_I(t)\rangle$,
which satisfies
$$i\hbar\frac{\partial}{\partial t} |\psi_I(t)\rangle = \hat{V}_I |\psi_I(t)\rangle
\,,$$
where $\hat{V}_I = \exp(i\hat{H}_0 t/\hbar) \hat{V}
\exp(-i\hat{H}_0 t/\hbar)$. Now that $\hat{H}_0$ has been
eliminated from the Schr\"{o}dinger equation, it seems that
dynamical phases due to the qubit free evolution would have no
effect. However, at the end of a calculation, physical
measurements have to be performed to read out the computational
results, and these measurements are generally performed in the lab
frame (the Schr\"odinger picture), in which the dynamical phases
reappear.  More specifically, the measurement of an observable
$\widehat{O}$ can be expressed as $\langle \psi_S(t)| \widehat{O}
|\psi_S(t)\rangle = \langle \psi_I(t)| \exp(i\hat{H}_0 t/\hbar)
\widehat{O} \exp(-i\hat{H}_0 t/\hbar) |\psi_I(t)\rangle = \langle
\psi_I(t)| \widehat{O}_I (t) |\psi_I(t)\rangle$. In other words,
if we prefer calculating the expectation value of a
time-independent operator, it has to be done in the
Schr\"{o}dinger picture. If
$|\psi_I(\tau)\rangle=\sum_j\alpha_j\,|j\,\rangle$ is the desired
final state, the Schr\"{o}dinger picture final state would take
the form
\begin{equation}
|\psi_S
(\tau)\rangle=\sum_j\,\alpha_j\,e^{-iE_j\tau}|j\,\rangle=\sum_j\,
\alpha_j\,e^{-i(L_j^{(1)}-L_j^{(0)})\tau\Delta}|j\,\rangle \,.
\end{equation}
Therefore, the phase-matching condition (\ref{eq-phase}) would
render the phases $\exp[-i(L_j^{(1)}-L_j^{(0)})\tau\Delta] = 1$,
so that it enforces the equivalence of interaction picture and
Schr\"odinger picture states, which ensures that the coherent
error arising from the free evolution during the delay can be
effectively eliminated.

In what follows, we illustrate our discussion with a few instances
of Shor's algorithm.

\subsection{An analytical example for factoring a small composite
number}

Let us first consider the factorization of the smallest composite
number $4$, which uses a two-qubit work register, a two-qubit
auxiliary register and $a=3$. After going through the four steps
of Shor's algorithm as discussed above, the final work register
state (Eq. (4)) is
\begin{widetext}
\begin{eqnarray}
|\psi(\tau+\tau_4)\rangle_W&=&
\frac{1}{\sqrt{2}}\left\{\frac{1}{\sqrt{2}}\left(|0_1\rangle_W+e^{-i\tau_4\Delta}|1_1\rangle_W\right)\otimes
\frac{1}{\sqrt{2}}\left[\zeta|0_0\rangle_W+\xi e^{-i\tau_4\Delta}|1_0\rangle_W\right]\right\}\nonumber\\
&=&\frac{1}{\sqrt{8}}\left[\zeta|0\rangle_W+\xi|1\rangle_W+e^{-i\tau_4\Delta}\zeta|2\rangle_W
+e^{-i\tau_4\Delta}\xi|3\rangle_W\right],
\end{eqnarray}
\end{widetext}
with $\zeta=1+e^{-i\tau\Delta}$, and $\xi=1-e^{-i\tau\Delta}$.
Here, $|\alpha_k\rangle_W$ refers to the logical states (with
$\alpha=0,1$) of the $k$th (with $k=0,1$) qubit in the work
register. In other hands,
$|0\rangle_W=|0_10_0\rangle_W,\,|1\rangle_W=|0_11_0\rangle,\,|2\rangle_W=|1_10_0\rangle$,
and $|3\rangle_W=|1_11_0\rangle_W$.

To derive Eq. (9), the measurement on the auxiliary register is
the projection $\hat{P}_A=|1\rangle_{A}\langle 1|_A$.
Measuring the work register in the computational basis, the state
(9) collapses to the expected one: either $|0\rangle_W$, or
$|2\rangle_W$, with probability
$p_e=|\zeta|^2=[1+\cos(\tau\Delta)]/4$. This implies that the
desired results ($|\zeta|^2=1/2$) are obtained, only if the
phase-matching condition (7) is satisfied. Equation (9) also shows
that the dynamical phase acquired by each qubit after the Fourier
transform does not result in any measurable physical effect.

\subsection{Numerical examples for factoring a few integers}

To quantitatively evaluate the effects of the dynamical phases
when running Shor's algorithm, we introduce two delay-dependent
functions: $p_e(k_e)$ is used to quantify the probability of
obtaining the correct result $k_e$, and
\begin{equation}
P_e=\sum_{k_e}p_e(k_e)
\end{equation}
is the probability of computing all the correct outputs. $P_e=1$
for an ideal computation process and for practical quantum
computers at the phase-matching time intervals consistent with
Eq.~(\ref{eq-phase}). For other delays not satisfying
Eq.~(\ref{eq-phase}) wrong results ($k\neq k_e$) can be obtained
so that $P_e<1$.

We now run the algorithm to factorize $N=21$ with $a=5$ using $9$
work qubits. Fig.~2 shows the various outputs and the
corresponding probabilities for different delay times $\tau$:
$\tau\Delta = 0,\, 0.4\pi,\, \pi, \,1.6\pi$, and $2\pi$.
\begin{figure}
\label{fig1}
\begin{center}
\vspace{-0.5cm}
\includegraphics[height=7.6cm,width=11.8cm]{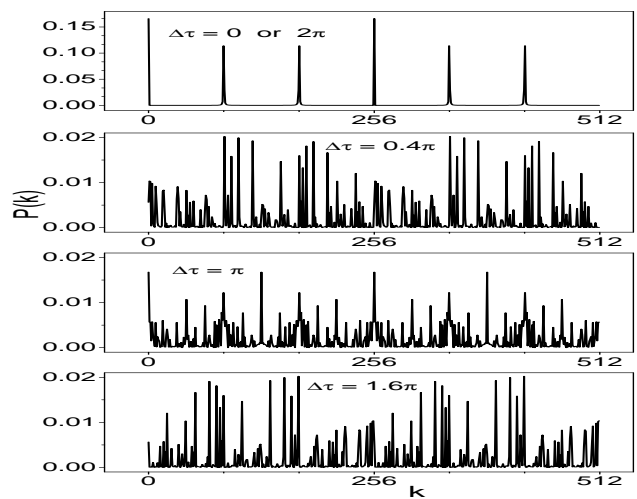}
\vspace{-1cm} \caption{The probability $P(k)$ (see Eq.~(6)) of
observing values of $k$ for different values of
$\tau\Delta=(\epsilon_1-\epsilon_0)\tau=0$,
$0.4\protect\pi$, $\protect\pi$, $%
1.6\protect\pi$, and $2\protect\pi$, given $N=21$, $q=512$, $a=5$,
and the expected order $r=6$. Here, $\tau$ is the total effective
delay time between unitary operations. The correct outputs are
obtained when the phase-matching condition
$\tau\Delta=2\protect\pi$ (or the ideal case $\tau\Delta=0$) is
satisfied. The probabilities of obtaining the correct outputs far
from the phase-matching conditions are very low (see the second,
third, and fourth panels. Note the different scales for the
vertical axes). Indeed, as shown in the bottom three panels, many
incorrect results are produced when the phase matching condition
given by Eq.~(\ref{eq-phase}) is not enforced.}
\end{center}
\end{figure}
It is seen from Fig.~2 that, when the phase-matching condition
(\ref{eq-phase}) is satisfied, the computed results are identical
to that of an ideal computation process with $\tau\Delta=0$.
Note in Fig.~2 that the maximum value of $P(k)\approx 0.2$ at the
matching condition and $P(k)< 0.02$ away from it.

We plot the delay-dependent $P_e$ in Fig.~3 for several examples:
factorizing $N=15$, $21$, and $33$, with $a=13$, $5$, and $5$, and
when using $4$, $9$, and $11$ work qubits, respectively. As is
shown in Fig.~3, the correct results are always obtained at the
phase-matching time intervals given by Eq.~(7). For other delay
cases, especially near the delay points satisfying the condition:
$\tau\,\Delta=(\epsilon_1-\epsilon_0)\,\tau=(2n-1)\pi$, the
correct results cannot be obtained (for the case where the
expected order is a power of two; see, e.g., the continuous line
for $r=4$ in Fig.~3) or may be obtained with very low
probabilities $P_e$ (for the cases where the order $r$ cannot
divide the given $q$ exactly; see, e.g., the lines for $r=6$, $10$
in Fig.~3). Of course, the dynamical oscillations can also be
suppressed by trivially setting up individual delays $\tau_m$ as
$\tau\Delta_m=2n\pi$. The key observation here is that
\textit{only the total delay time}, instead of the duration for
every delay, \textit{needs to be set up accurately to avoid the
coherent dynamical phase error}.

Classically, higher precision is usually obtained by using more
computational bits. However, this is not necessarily the case in
practical quantum computation. Indeed, in the current example of
Shor's algorithm, after taking into consideration the influence of
the time delays between consecutive computational operations, the
more qubits are used, the {\it lower} the computational efficiency
is. This relationship is clearly demonstrated in Fig.~4, which
shows that {\it the probability of obtaining any one of the
correct results decreases exponentially when increasing the number
of qubits of the work register}. Such a scenario is to be
expected, since the number of possible outputs in the final
measurement increases exponentially with the number of the work
qubits, which makes the constructive interference in Eq.~(5) for
the probability $P(k)$ harder to achieve if $\tau\Delta$ deviates
from the phase-matching condition (7). At the exact points when
$(\epsilon_1-\epsilon_0)\tau=2n\pi$, the constructive interference
of the superposition wave functions ensures that the computational
accuracy is independent of the number of involved qubits.
\begin{figure}[tbp]
\begin{center}
\vspace{2cm}
\includegraphics[height=6.6cm,width=13.2cm]{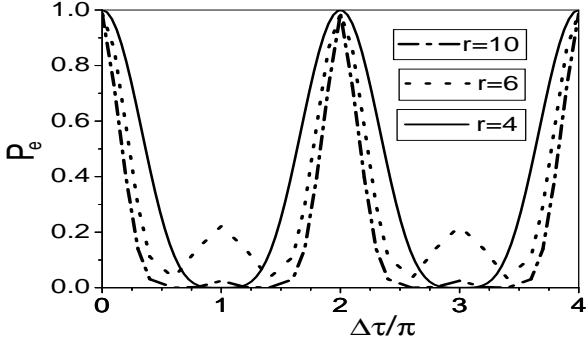}
\vspace{-4.5cm}
\end{center}
\caption{The probability $P_e$ of obtaining the correct results versus
$\Delta%
\protect\tau=(\epsilon_1-\epsilon_0)\tau$ for running Shor's
factoring algorithm in the presence of
delays. The lines for $r=4$, $6$, $10$ correspond to the cases where $4$, $9$%
, $11$ work qubits, given $q=16$, $512$, $2048$, are used to factorize $N=15$%
, $21$, $33$ with $a=13$, $5$, $5$, respectively. Note that the expected
outputs can be obtained at phase-matching points: $\Delta\protect\tau=2%
\protect\pi,4\protect\pi$.} \label{fig3}
\end{figure}
\begin{figure}[tbp]
\begin{center}
\vspace{2cm}
\includegraphics[height=6cm,width=12cm]
{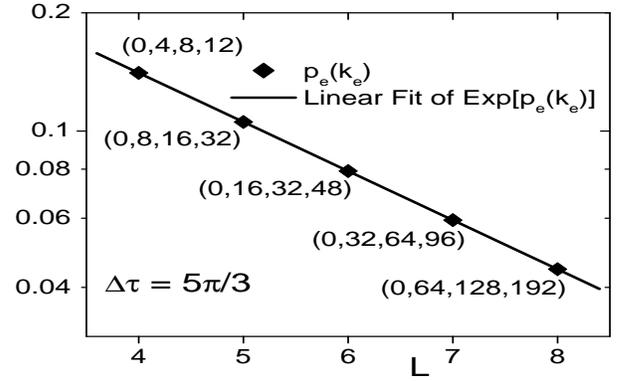} \vspace{-4cm}
\end{center}
\caption{The probability $p_e(k_e)$ of obtaining one of the
correct results versus the number $L$ of work qubits used to run
the quantum algorithm factorizing $N=15$ in the presence of a
delay
$\Delta\protect\tau=(\epsilon_1-\epsilon_0)\tau=5\protect\pi/3$.
The straight line shows that this probability $p_e(k_e)$ decreases
exponentially with the number $L$ of qubits used. The points on
the line show the probability of obtaining one of
the correct outputs $k_e=(0$, $4$, $8$, $12)$ for 4-qubits, $(0$, $8$, $16$%
, $24)$ for 5-qubits, $(0$, $16$, $32$, $48)$ for 6-qubits, $(0$ , $32$, $64$%
, $96)$ for 7-qubits, and $(0$, $64$, $128$, $192)$ for 8-qubits
cases, respectively.} \label{fig4}
\end{figure}
\begin{figure}[tbp]
\begin{center}
\vspace{2cm}
\includegraphics[height=6cm,width=12cm]
{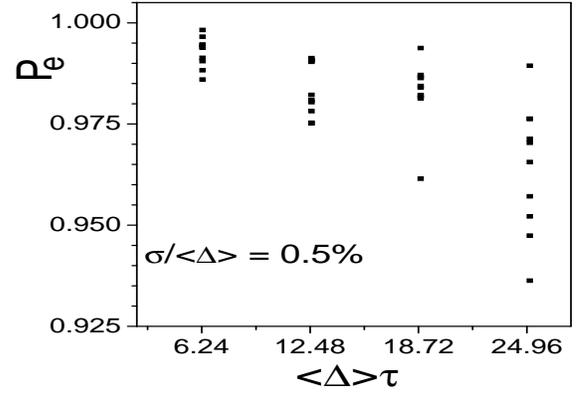} \vspace{-3.8cm}
\end{center}
\caption{The probabilities $P_e$ (for factorizing $N=15$ using $8$
work qubits) of obtaining the correct results for different
phase-matching
cases: $\langle\Delta\rangle\protect\tau=2\protect\pi$, $4\protect\pi$, $6%
\protect\pi$, $8\protect\pi$, with a common Gaussian energy
splitting fluctuation with
$\protect\sigma/\langle\Delta\rangle=0.5\%$. Note that this
probability $P_e$ is higher at the phase matching points with
shorter total delay time $\tau$.} \label{fig5}
\end{figure}
\begin{figure}[tbp]
\begin{center}
\vspace{2cm}
\includegraphics[height=6cm,width=12cm] {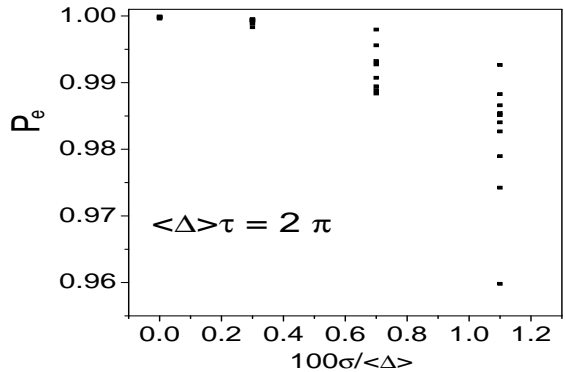}
\vspace{-4cm}
\end{center}
\caption{The probabilities $P_e$ (for factorizing $N=15$ using $8$
work qubits) of obtaining the correct results for different
fluctuations
of energy splittings: $\protect\sigma/\langle\Delta\rangle=0.01\%,0.3\%,0.7%
\%,1.1\%$, with a common phase-matching point: $\langle\Delta\rangle\protect%
\tau=2\protect\pi$. Note that the probability at the pase-matching
point is still sufficiently high, even if the energy splittings of
the qubits exist with certain fluctuations around the average
value $\langle\Delta\rangle$.} \label{fig6}
\end{figure}

\subsection{Effect of energy splitting inhomogeneity}

In the calculations up to now, we have assumed that all qubits
possess an identical energy splitting
$\Delta=\epsilon_1-\epsilon_0$. In reality, especially for the
solid state quantum systems such as the Josephson junction qubits
and quantum dot trapped spins, different qubits may have slightly
{\it different} energy splittings due to system inhomogeneity. The
logical states with the same energy in the ``identical qubit''
assumption (e.g., $|1_30_20_10_0\rangle$ and
$|0_30_20_11_0\rangle$) may now have slightly different energies.
A critical question then is how robust the phase matching
condition (\ref{eq-phase}) is for a system of multiple qubits with
fluctuations in the qubit energy splittings. Here we provide
quantitative answers to this important question by numerically
simulating Shor's algorithm assuming a Gaussian distribution for
the qubit energy splittings. In other words, the energy splitting
$\Delta_j$ of the $j$th qubit is chosen randomly according to the
distribution function
\begin{equation}
P(\Delta_j)=\frac{1}{\sqrt{2\pi}\,\sigma} \exp
\left[-\,\frac{(\Delta_j - \langle \Delta \rangle)^2}{2\sigma^2}
\right]
\end{equation}
around an average value $\langle\Delta\rangle$ and width $\sigma$.
Figure 5 shows that the probability of obtaining correct answers
decreases as the total time delay $\tau$ increases.  Also, Fig.~6
shows the dependence of $P_e$ on the width of the qubit energy
splitting distribution $\sigma$, with the delay condition set at
$\langle\Delta\rangle\tau = 2\pi$. As expected, a quantum computer
runs with higher efficiencies for shorter time delays $\tau$ and
for narrower distributions $P(\Delta_j)$ of energy splittings. In
essence, here we study an effect similar to inhomogeneous
broadening, which is not a true dephasing effect. This is
consistent with our focus in this paper on the coherent errors
instead of the incoherent ones.

\section{Conclusions and Discussions}
When a real quantum computer performs a computational task, there
must be unavoidable time intervals between consecutive unitary
operations. During these delays, the wavefunction of a system with
non-zero free Hamiltonian would acquire relative dynamical phases,
if the two states for each qubit have different energies.  These
dynamical phases lead to fast oscillations in the total
wavefunction, and modify the desired quantum interference required
by quantum algorithms, which in turn reduce the probability of
obtaining correct computational results.

Here we have studied the effects of the dynamical phases in
running a quantum algorithm (more specifically, Shor's factoring
algorithm).  We point out that a phase-matching condition can
potentially help allieviate the interference problems caused by
the dynamical phases, and this condition is closely related to
establishing the equivalence between quantum states in the
Schr\"{o}dinger picture and the interaction picture through a
quantum computation process. In the presence of coherent phase
errors, we have demonstrated that, the probability of obtaining
the correct answer decreases exponentially with increasing number
of qubits of the work register. In addition, Shor's algorithm
fails for the worst case scenario of $\tau\Delta=(2n-1)\pi$ if the
expected order $r$ is a power of two.  We have further shown that
the phase-matching condition studied here is quite robust in the
presence of small fluctuations in the qubit energy splittings.
Unlike the refocusing technique in NMR experiments
\cite{Vandersypen01}, which deals with unwanted evolutions due to
uncontrolled qubit interaction, we have shown here that by
properly setting the {\it total} effective delay, the unwanted
oscillations of the superposed wavefunctions due to the free
Hamiltonians of the bare qubits can be effectively suppressed,
thus the desired output can be obtained without additional
operations. This implies that the quantum computing may be
performed in an effective interaction picture, in which coherent
errors arising from the free evolution of the bare qubits during
the operational delay can be automatically avoided.

We emphasize that the present simplified approach only treats the
delays between two sequential functional operations and neglects
those inside these transforms. In fact, each functional transform,
which is actually equivalent to a multi-qubit gate, can be, in
principle, implemented exactly by using only one-time evolution
\cite{wei04-1,Niskanen03}. This ``coarse-grained" one-step
implementation implies that the evolutions relating to the various
parts of the total Hamiltonian have been well controlled.
Therefore, the operational delays, relating only to the free
evolution ruled by the free Hamiltonian of the bare physical
qubits, within each one of these larger functional building blocks
are assumed to be zero. Also, the dynamical phases acquired by the
superposed wavefunctions can be added up for the operational
delays before and after each functional transformation. Therefore,
the phase-matching condition (7) exists for the {\it total} delay.

The present calculation is done assuming that Shor's algorithm is
accomplished in 5 lumped steps. A simple analysis can prove that,
even if using an actual elementary gate array model, e.g., shown
in figure 7 (for implementing the initializations by using the
Hadamard gates and the quantum Fourier transformation),
\begin{figure}[tbp]
\begin{center}
\vspace{2cm}
\includegraphics[height=10.2cm,width=11.2cm] {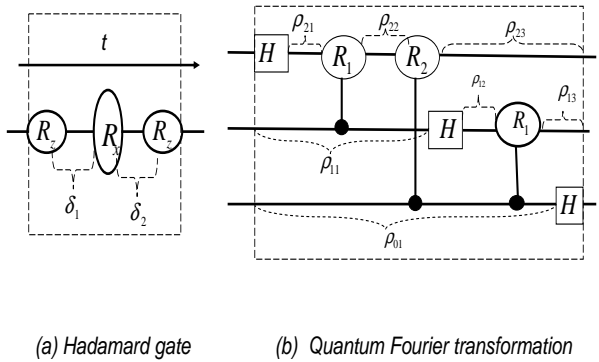}
\vspace{-7.6cm}
\end{center}
\caption{Quantum circuits formed by the elementary single- and
two-qubit logic gates for performing (a) Hadamard gate for one
qubit, and (b) quantum Fourier transformation for three-qubit.
Here, $\delta_l\,(l=1,2,...)$ and $\rho_{kl}\,(k=0,1,2,...)$ refer
to the operational delays inside them, respectively. In the
logical basis, the single-qubit gate
$\hat{R}_z=\exp(i\pi\sigma_z/4)$ and the two-qubit
controlled-phase gate $R_k=|00\rangle\langle 00|+|01\rangle\langle
01|+|10\rangle\langle 10|+\exp(2i\pi/2^k)|11\rangle\langle 11|$
are diagonal, while the single-qubit
$\hat{R}_x=\exp(i\pi\sigma_x/4)$ is not.}
\end{figure}
the proposed phase-matching conditions (in terms of the total
delay time instead of individual delay times of each operational
delay) for avoiding the coherent phase errors are still valid. The
key is that, only two elementary non-diagonal operations (i.e.,
$\hat{R}_x$ in Hadamard gates) are applied to each qubit in the
work register (see Fig. 7). The qubit is in a product state before
the first non-diagonal $\hat{R}_x$ gate, while the delays after
the second non-diagonal $\hat{R}_x$ in the corresponding Hadamard
gate do not affect the results of projective measurement (see,
e.g., Eq. (9)). Therefore, the dynamical phases acquired in
different effective operational delays accumulate even when the
operational delays inside the functional steps are considered.

In the present approach, we have assumed that every qubit in the
work register has the same waiting time $\tau_j^+$ for each
effective operational delay. In practice, this assumption is not
necessary. Indeed, in the elementary gate array model, the waiting
times for different qubits would have been different. However, the
phase-matching condition Eq. (7) needs only a slight modification
in this case, so that it becomes a condition for each
qubit~\cite{Wei04-2}:
$\tau_k\Delta_k=2n_k\pi,k=1,2,...;n_k=1,2,..$ for each qubit.
Here, $\Delta_k$ and $\tau_k$ are the energy splitting and {\it
total} controllable effective delay of the $k$th qubit in the work
register, respectively.

Finally, we emphasize that the results presented in this paper,
through obtained using a simple model for the delays, clearly
demonstrate the necessity of taking into consideration the
dynamical phases of the qubits in implementing quantum algorithms.

\section*{Acknowledgments}
This work was supported in part by the National Security Agency
(NSA) and Advanced Research and Development Activity (ARDA) under
Air Force Office of Research (AFOSR) contract number
F49620-02-1-0334, and by the National Science Foundation grant No.
EIA-0130383.

\end{document}